\documentclass[%
preprint,
 amsmath,amssymb,
 aps,
]{revtex4-2}

\usepackage{graphicx}
\usepackage{dcolumn}
\usepackage{bm}

\begin{document}


\title{Comment on:"Investigating the time dependence of neutron-proton equilibration using molecular
dynamics simulations" by A.Jedele \textit{et al}: Phys Rev C 107, 024601 (2023)}\

\author{Massimo Papa}
 \email{Massimo.Papa@ct.infn.it}
\affiliation{%
 Istituto Nazionale Fisica Nucleare-Sezione di Catania, Corso Italia 57, I-95129 Catania, Italy\\
}%


%

\date{\today}

\begin{abstract}
In this paper the authors discuss neutron-proton equilibration 
 process induced on the $^{70}Zn + ^{70}Zn$ system at 35 MeV/nucleon comparing experimental results with the Anti-symmetrized Molecular Dynamics and the Constrained Molecular Dynamics model simulations (COMD) . The comment contains observations on the improper use of the COMD model including some misinterpretations of the obtained results.
\end{abstract}

\maketitle
\section{\label{sec:level1}Introduction}
The Molecular Dynamics Models (MD) represent a wide class of semi-classical models able to describe
the main feature of the dynamics associated to the Heavy Ion Collision (HIC) around the
Fermi energy and beyond (see Ref.\cite{wolt} for a review).
The semi-classical approximation relies in the choice to represent the single  particle wave functions by means of wave-packets (WP) with fixed widths. The equation of motion for the WP centers are determined from the Hamiltonian obtained through the convolution
of an effective microscopic interaction with the WP. The N-body wave functions is supposed to be the  direct product of the single particle wave functions.
With a reasonable computing time, this kind of approaches are able to follow the HIC dynamics
for several hundreds of fm/c on different thousands of realizations. Dynamical fluctuations are  spontaneously produced leading to the formation of cluster because of the localization effects associated to the WP.
In particular, the Constrained Molecular Dynamics approach, as in the old original version COMD\cite{comd} , is characterized by the introduction of a numerical  procedure 
with which the equations of motion of the WP centres can be solved fulfilling the Pauli Principle in phase-space at each time step.
 With this innovation the studied nuclear many-body system acquire the characterizing properties of a Fermionic system.
The constraint has important effects on  dynamics of the system: it allows a reliable evaluation of the Pauli blocking factors associated to the nucleon-nucleon collision
process (residual repulsive interaction) for the system of Fermions. 
 Moreover, a cooling-warming procedure\cite{comd,jcomp} coupled with the constraint allows
to produce long time stable initial configurations (GS "ground state" configurations) which are determining for a reliable description of the subsequent collision process. 
This implies the implementation of a fundamental  preliminary phase to the study of the dynamics of colliding systems in which the initial conditions are appropriately prepared.
(see also next section).
Obviously for MD approaches the intrinsically semi-classical treatment due to the use of stable WP as single particle wave function and to the use of relatively simple effective interactions represents stringent
limits for a full description of the dynamical evolution. 
In the dynamics of collision processes, after the phase of fast emission  of nucleons and light fragments (typically in a few hundred of fm/c) and after a possible delayed "break-up" phase (typically within 1000 fm/c) of the primary sources, the subsequent cooling process of the hot big fragments (see next section)  is usually achieved through quantum-statistical decay models.
Another important limit relies on the structural description of light particle and fragments which as it is well known is instead dominated by quantum effects.
In the following section comments related to the paper under discussion  are presented. They mainly concerns
the proper use of the  COMD code in relation to the above illustrated features and 
some misinterpretations of the obtained results.

\section{\label{sec:level2}Comments}
-1) The old original version of the  Constrained Molecular 
Dynamics model (named COMD), as others MD models, was not designed to describe details of the production of  light clusters. This on the other hand  was also indicated in Ref.\cite{comd}(see for example the abstract). 

-2) In these calculations is not appropriate  to apply the statistical decay codes like Gemini for the de-excitation  of the light nuclei with charge Z=6-$\simeq$10 produced by the COMD because for such light nuclei the GS binding produced by the model can not be unambiguously defined and  does not fit the experimental values. This just because this semi-classical approach cannot  describe relevant quantum effects of the light particles/clusters.  
Rather, it is appropriate to accept them as obtained at the end of the dynamical evolution.  They are in fact formed spontaneously  and  survive up to the long time $t_{M}$ (time interval of the dynamical evolution). They are therefore sufficiently stable to be thought of as the representative of a kind of “ground state” (GS) consistent with the semi-classical model. 
For the more heavy fragments the situation is different because they instead  can keep easily excitation energy for longer times.

Other aspects concern the time. 

-3) In that work  COMD calculations are performed by evolving the dynamics up to $t_{M}$ 1000 fm/c and they are compared with the Anti-symmetrized Molecular Dynamics ones at 300 fm/c. But for COMD no specification of the stability time and of the adopted procedure used to get the initial configurations is given. 
Within the COMD model, the selection of sufficiently stable initial configurations is fundamental to obtain a reliable dynamics.
For each nucleus  good
“ground state” configurations (GS) are obtained after a cooling-warming procedure coupled with the constraint\cite{comd,jcomp} . For each system good GS 
 configurations (with an average binding tipycally of about 8-8.7 MeV/nucelon)can be selected by means of fine adjustments involving the nuclear radii R (within some percent of the experimental values) and of the surface energy  corrective term $E_{sup}$.  Then R and $E_{sup}$ can be selected in a region corresponding to the maximum in the average stability time $<T_{stab}>$ evaluated on many attempted configurations.  $T_{stab}$ of the selected configurations should be at least of the order of different hundreds of fm/c.

-4) In this work it is written: “The larger statistical contribution in the COMD distribution is due to the clusterization parameters. Nucleons are deﬁned to be in the same fragment if the center of mass of the nucleons is within 2.76 fm of each other. A large number of events still have fragments within the clusterization radius at the end of the COMD simulation and are therefore deﬁned as one fragment.”  

This statement  deserves different remarks.

-a) To avoid the problem mentioned in the previous point or to test the interpretation of the results obtained at 1000 fm/c one can increase $t_{M}$ waiting for a definite separation. $t_{M}$ is not a fixed parameter of the model. $t_{M}$ should be chosen depending on what one want to get from the calculations, and what information produced by the model the authors judge more or less confident  (fast pre-equilibrium, including or not, more delayed emissions)\cite{DF} . 

-b) In any case, i.e. also  with an high rate of fragmentation at 1000fm/c as declared in the paper (never noticed with high rate by the writer) after this long time no memory of the initial orientation will be present and this will not destroy the isotropy. The average equilibrium deformations after the  separation will be also  small.

-c) At the  distance $d$= 2.76 fm (in my current version  $d$= 2.56 fm but this is not critical, also $d$ is not a parameter of the COMD model) the particles are still in interaction according to their Gaussian widths $\sigma_{r}$; (for $d$= 2.76 fm the interaction is larger than 0.2 the maximum value for a complete overlap) i.e. they still  feel the nuclear interaction trough the Gaussian tails.
From a simple liquid drop model one obtain at the saturation density $\rho_{0}$: $d=2r_{0}=2(\dfrac{3}{4\pi\rho_{0}})^{1/3}=2.26$ fm.
 Therefore this choice is quite reasonable. The evaluation of $d$ can take into account also the fact that excited big fragments have usually a lower density with respect the saturation one, but in any case the choice of $d$, within the above criteria,  never becomes critical if a reasonable choice of $t_{M}$ is done.

\end{document}